\documentclass[12pt]{article}
\usepackage{graphicx}
\newsavebox{\sboxpubnumber}
\newsavebox{\sboxpubdate}

\newcommand{\pubnumber}[1]{\begin{lrbox}{\sboxpubnumber}{\begin{tabular}{l} #1 \\
				 \usebox{\sboxpubdate}
				 \end{tabular}}
                           \end{lrbox}
                           \pubblock}
\newcommand{\Title}[1]{\begin{center} {\Large #1 } \end{center}}
\newcommand{\Author}[1]{\begin{center}{ \sc #1} \end{center}}
\newcommand{\Address}[1]{\begin{center}{ \it #1} \end{center}}

\newcommand{\pubblock}{\rightline{
			\usebox{\sboxpubnumber}}}
\newenvironment{Abstract}{\begin{quotation}  }{\end{quotation}}
\newenvironment{Presented}{\begin{quotation} \begin{center}
             PRESENTED AT\end{center}\bigskip
      \begin{center}\begin{large}}{\end{large}\end{center}
      \end{quotation}}
\newcommand{\Acknowledgements}{\bigskip  \bigskip \begin{center} 
\begin{large}
             \bf ACKNOWLEDGEMENTS \end{large}\end{center}}
\def\vf{\varphi}
\def\laq{\raise 0.4ex\hbox{$<$}\kern -0.8em\lower 0.62 ex\hbox{$\sim$}}
\def\gaq{\raise 0.4ex\hbox{$>$}\kern -0.7em\lower 0.62 ex\hbox{$\sim$}}

\begin{document}
\begin{titlepage}
\pubnumber{UNIL-IPT 01-16 \\ YYY-YYYYYY} 
\vfill
{}\Title{Cosmological Magnetogenesis:\\ what we know and what we would like to 
know}
\vfill
\Author{Massimo Giovannini\footnote{Electronic address: 
Massimo.Giovannini@ipt.unil.ch}}
\Address{Institute of Theoretical Physics, University of Lausanne, \\
BSP CH-1015, Dorigny, Switzerland}
\vfill
\begin{Abstract}
Problems and perspectives concerning  the existence 
of large-scale magnetic fields are described. Heeding observations,
possible origins and implications of magnetic fields in spiral 
galaxies and in regular clusters are 
scrutinized in different cosmological frameworks including 
scenarios where conformal invariance is broken 
because of the evolution of the gauge couplings. The generation 
of the BAU via hypermagnetic knots is  reviewed. It is also argued
that stochastic GW backgrounds can be generated in the LISA frequency range 
thanks to the presence of hypermagnetic fields.
\end{Abstract}
\vfill
\begin{Presented}
    COSMO-01 \\
    Rovaniemi, Finland, \\
    August 29 -- September 4, 2001
\end{Presented}
\vfill
\end{titlepage}
\def\thefootnote{\fnsymbol{footnote}}
\setcounter{footnote}{0}
\renewcommand{\theequation}{1.\arabic{equation}}
\setcounter{equation}{0} 
\section{Old stories and new measurements}
Enrico Fermi was born hundred yers ago, on September 29 1901. 
I then take the chance of starting my contribution 
to the COSMO-01 meeting by reminding that the first speculations concerning 
the existence of large scale magnetic fields are due to Fermi who published 
a single author paper in the issue 75 of Physical Review \cite{fermi}. 
It was 1949.
At that time a debate was going on in the community. On one side   
Alfv\'en  \cite{alv1,alv2} and, simultaneously,
Richtmyer and Teller \cite{alv3}  
were claiming that high energy cosmic rays are in 
equilibrium with stars 
\cite{alv1,alv2}. 
In a different perspective
Fermi gave arguments supporting the idea 
that high energy cosmic rays are in equilibrium 
with the galaxy. The idea of Fermi was that that cosmic 
rays are a {\em global } galactic phenomenon. 
The suggestion of Alv\'en was that 
cosmic rays are a {\em local} solar phenomenon. 
In order to make his argument consistent Fermi needed a large scale (galactic)
magnetic field as large as the $\mu$ Gauss over the size of the galaxy.  
Four years later, Fermi and Chandrasekar \cite{fermi2} developed the first 
theory of the gravitational instability in the presence 
of large scale magnetic fields : an ancestor of what 
we call today dynamo theory. 

Fifty years after Fermi's speculations, large scale magnetic fields represent 
an intriguing triple point where cosmic ray physics, cosmology and 
astrophysics meet for different (but related) purposes. 
Today large scale magnetic fields are measured not only 
in the Milky Way but also in other members of the local group and 
even in regular Abell clusters.

From the experimental point of view the best studied field is the 
one of the Milky Way where various observational techniques can 
be exploited. In particular Zeeman splitting estimates offer 
a good tool in order to study magnetic fields which are locally strong, like,
for instance, the ones present in compact OH  sources. Unfortunately, due
to the well known limitations of Doppler broadening, Zeeman splitting 
estimates fail for large portions of the interstellar medium where 
the hyperfine splitting of neutral hydrogen would represent 
an ideal line in order to infer the strength of the magnetic field.
Faraday rotation measurements, combined with synchrotron emission, 
represent probably the best option in order to measure large scale 
magnetic fields even outside our galaxy. 

The known limitation of Faraday rotation stems from the 
need of an independent measure of the electron density along the line of 
sight: today there is no clear experimental evidence of why the magnetic 
field right outside the galaxy should be orders of magnitude  smaller 
than the $\mu$ Gauss ( n Gauss, as often speculated). 

Since various  models of large scale magnetic field generation 
predict the existence of magnetic fields not only 
in galaxies but also inside clusters, it would be interesting 
to know if magnetic fields are present inside regular Abell clusters. 
Various results in this direction have been reported 
\cite{cl1,cl2,cl3,cl4}. Some studies during the past
decade \cite{cl1,cl2}
dealt mainly with the case of a single cluster (more specifically the 
Coma cluster). Many radio sources inside the
cluster were targeted with Faraday rotation 
measurements (RM). The study of many radio sources inside 
different clusters presents experimental 
problems due to the sensitivity limitations of 
radio-astronomical facilities. Hence the strategy 
is currently to study a sample of clusters each with one or two 
bright radio-sources inside.

In the past it was shown that regular clusters have a cores with
a detectable component of RM \cite{cl3,cl4}. Recent results 
suggest that $\mu$ Gauss magnetic fields are indeed detected 
inside regular clusters \cite{cl5}. Inside the cluster means 
in the inter-cluster medium. Therefore, these magnetic fields
cannot be associated with individual galaxies.
 
Regular Abell clusters with strong x-ray emission were 
studied using a twofold technique \cite{cl5,cl6}. From the ROSAT 
\footnote{The ROetgen SATellite was flying from June 1991 to February 1999.
ROSAT provided a map of the x-ray sky in the range $0.1$--$2.5$ keV.}
full sky survey the electron density has been determined.
Faraday RM (for the same set 
of 16 Abell clusters) has been estimated through observations at the VLA 
\footnote{The Very Large Array
telescope is a radio-astronomical facility consisting 
of 27 parabolic antennas spread around 20 km in the New Mexico desert.}.
The amusing result (confirming previous claims based only on one cluster 
\cite{cl1,cl2}) is that x-ray bright Abell clusters
 possess a magnetic field of $\mu$ Gauss 
strength.The clusters have been selected in order 
to show similar morphological features. All the 16 clusters 
monitored with this technique are at low red-shift ($z<0.1$) 
and at high galactic latitude ($|b|>20^{0}$).

These recent developments are rather promising 
and establish a clear connection between radio-astronomical 
techniques  and the improvements in the knowledge 
of  x-ray sky. There are 
various satellite missions mapping the
x-ray sky at low energies (ASCA, CHANDRA, NEWTON 
\footnote{ASCA is operating between $0.4$ AND $10$ keV and it is 
flying since February 1993. CHANDRA (NASA mission)  and NEWTON (ESA mission) 
have an energy range 
comparable with the one of ASCA and were launched, almost 
simultaneously, in 1999.}). There is 
the hope that a more precise knowledge of the surface brightness of regular
clusters will help in the experimental determination of large scale 
magnetic fields between galaxies.

The present paper is organized as follows. In Section II 
the main aspects of the amplification of large scale magnetic fields 
in the galaxy will be briefly outlined in a critical way. 
In Section III the attention will be concentrated 
on the general features of cosmological magnetogenesis. It will be argued 
that causal and inflationary mechanisms are not alaternative but complementary.
In Section IV and V the interplay between the evolution 
of the gauge couplings and the generation of large scale 
magnetic fields will be reviewed. Section IV is devoted to 
the case of dynamical extra-dimensions leading to an effective 
evolution of the gauge 
couplings. Section V deals with a model where gauge couplings evolve 
in four dimensions and in a standard framework of cosmological
evolution. Section VI addresses the possible effects of primordial 
magnetic fields 
on the polarization of the Cosmic Microwave Background 
(CMB). In Section VII it will be shown that 
the existence 
of primordial magnetic fields can be used in order to generate the 
baryon asymmetry of the Universe (BAU). At the same time, 
the hypermagnetic fields present at the electroweak epoch 
can lead to a gravitational waves background  not only 
in the LISA frequency range, but also at higher frequencies.

\renewcommand{\theequation}{2.\arabic{equation}}
\setcounter{equation}{0} 
\section{The galaxy as a gravitationally bound tokamak}

To address the problem of the origin of large scale magnetic fields means to 
write down the  equations describing their evolution.  This 
aspect is relatively well understood at least for what concerns the late 
stages of the evolution of the galaxy. A number of excellent 
textbooks and reviews can be consulted \cite{rev1,rev2,rev3,rev4,kul}. 
Here only few important points will be made clear.

As far as the evolution of 
magnetic fields is concerned, the 
galaxy is a gravitationally bound system formed by fluid of charged 
particles  which is globally 
neutral for scales larger than the Debye sphere.
In the interstellar medium, where the electron density 
is approximately $3\times 10^{-2} ~ {\rm cm}^{-3}$ (as it 
can be estimated from the dispersion of pulsar signals) the 
Debye sphere has a radius of roughly $10$ m.
 
Moreover, the galaxy
is rotating with a typical rotation period of $3\times10^{8}$ yrs. 
The evolution equations of this system are, 
physically, the same equations describing the dynamics 
of electromagnetic fields inside a tokamak. As in the case 
of a tokamak we have two choices. We can study the full kinetic system 
(the Vlasov-landau equations \cite{vla,lan}) 
or we can rely on the magnetohydrodynamical (MHD)
treatment. 
Already in flat space \cite{lif}, and, a fortiori, in curved space \cite{ber}, 
the kinetic approach is important once we deal with 
electric fields dissipation, charge and current density 
fluctuations and, in more general terms, with all the high 
frequency and small length scale phenomena in the plasma \cite{krall,bis}.

Consider a conformally flat Friedmann-Robertson-Walker (FRW) 
metric written using the  conformal time coordinate 
\begin{equation}
ds^2 = a^2(\eta)[d\eta^2 - d\vec{x}^2].
\label{metric1}
\end{equation}
Furthermore, consider an equilibrium homogeneous and
isotropic conducting plasma, characterized by a distribution function
$f_0(p)$ common for both positively and negatively charged
ultrarelativistic particles (for example, electrons and positrons) .
Suppose now that this plasma is slightly perturbed, so that the
distribution functions are
\begin{equation}
f_{+}(\vec{x}, \vec{p}, \eta) = f_0(p) + 
\delta f_{+}(\vec{x},\vec{p},\eta),
\,\,\,\,\,\,\,\,f_{-}(\vec{x},\vec{p}, \eta) =  
f_0(p) + \delta f_{-}(\vec{x},\vec{p},\eta),
\end{equation}
where $+$ refers to positrons and $-$ to electrons, and $\vec{p}$ is
the conformal momentum.  The Vlasov equation   defining the
curved-space evolution of the perturbed distributions can  be written
as  \cite{mg2}
\begin{eqnarray} 
&&\frac{\partial f_{+}}{\partial \eta} + 
\vec{v} \cdot \frac{\partial f_{+}}{\partial\vec{x}} + e ( \vec{E} + 
\vec{v}\times \vec{B}) \cdot\frac{\partial f_{+}}{\partial \vec{p}} = 
\biggl( 
\frac{\partial f_{+}}{\partial \eta}\biggr)_{\rm coll}
\label{Vl+},\\
&&\frac{\partial f_{-}}{\partial \eta} + 
\vec{v} \cdot \frac{\partial f_{-}}{\partial\vec{x}} - e ( \vec{E} + 
\vec{v}\times \vec{B})\cdot \frac{\partial f_{-}}{\partial \vec{p}} = 
\biggl(\frac{\partial f_{-}}{\partial\eta}\biggr)_{\rm coll}
\label{Vl-},
\end{eqnarray}
where the two terms appearing at the right hand side of each 
equation are the collision terms. The electric and 
magnetic fields are rescaled by the second power of the scale factor. 
 This system of equation represents 
the curved space extension of the Vlasov-Landau approach  to plasma
fluctuations \cite{vla,lan}. All particle number densities here are
related to the comoving volume.
By subtracting Eqs. (\ref{Vl+}) and (\ref{Vl-})  we obtain the
equations relating the fluctuations of  the distributions functions
of the charged particles present in the plasma  to the induced gauge
field fluctuations:
\begin{eqnarray}
&&\frac{\partial}{\partial\eta} f(\vec{x}, \vec{p},t) + 
\vec{v}\cdot \frac{\partial }{\partial\vec{x}} f(\vec{x},\vec{p},t) 
+ 2 e \vec{E}\cdot 
\frac{\partial f_0}{\partial \vec{p}} =- \nu(p) f,
\nonumber\\
&& \vec{\nabla} \cdot \vec{E} = e \int d^3 p f(\vec{x},\vec{p},\eta),
\nonumber\\
&& \vec{\nabla}\times \vec{E} + \vec{B}' =0,
\nonumber\\
&& \vec{\nabla}\cdot \vec{B} =0,
\nonumber\\
&& \vec{\nabla}\times \vec{B} -\vec{E}'= \int d^3 p \vec{v} 
f(\vec{x}, \vec{p},\eta),
\label{Vlasov}
\end{eqnarray}
where $f(\vec{x}, \vec{p}, \eta) = \delta f_{+}(\vec{x},\vec{p},\eta)
-  \delta f_{-}(\vec{x},\vec{p},\eta)$ and $\nu(p)$ is a typical
frequency of collisions \cite{ber}. 

Now, if $\delta f_{\pm}(\vec{x},\vec{p},\eta)\neq 0$ at the beginning of the
radiation dominated epoch $\eta_0$  and
$E(\vec{x},\eta_0)\simeq B(\vec{x},\eta_0)=0$ initially, the magnetic
field at later times can be found from Eqs. (\ref{Vlasov}) \cite{lif}.
Various useful generalizations of the Vlasov-Landau system to curved spaces 
is given in \cite{vlcur1,vlcur2,vlcur3}.

For scales sufficiently large compared with the Debye sphere and for 
frequencies sufficiently small compared with the plasma frequency the
spectrum of plasma excitations obtained from the kinetic 
theory matches the  MHD spectrum. 

Since the galaxy is rotating and since the conditions 
of validity of the MHD approximation
are met, it is possible to use the so-called dynamo instability in order 
amplify a small magnetic inhomogeneity up to the 
observed value. A necessary condition in order to implement this 
idea is that the flow should not be mirror symmetric, i.e. 
$\langle \vec{v} \cdot \vec{\nabla} \times \vec{v} \rangle \neq 0$, where 
$\vec{v}$ is the bulk velocity of the plasma.
The suggestion that the mean magnetic field, in a randomly moving 
medium ( with non mirror-symmetric flow), 
could grow was first proposed 
by Parker \cite{park}.  For a recent (critical) 
review on galactic dynamos see \cite{kul}.
MHD equations can be derived from a microscopic (kinetic) 
approach and also from a macroscopic approach where 
the displacement current is neglected \cite{krall}. 
If the displacement 
current is neglected the electric field can be expressed 
using the Ohm law and the magnetic diffusivity equation is obtained
\begin{equation}
\frac{\partial \vec{B}}{\partial\eta} = \vec{\nabla} \times(\vec{v} 
\times \vec{B}) + \frac{1}{\sigma} 
\nabla^2 \vec{B}.
\label{mdiff}
\end{equation}
The conductivity $\sigma$ appearing in Eq. (\ref{mdiff})
is a global quantity which can be computed in a kinetic approach \cite{krall}
during a given phase of evolution of the background 
geometry \cite{mg2}.
The term containing the bulk velocity field is called dynamo term 
and it receives contribution provided parity is globally broken over the 
physical size of the plasma. In Eq. (\ref{mdiff}) 
the contribution containing the conductivity is usually called magnetic diffusivity term.

The ratio of the two terms on the r.h.s. of Eq. (\ref{mdiff})
 defines the  magnetic Reynolds number 
\begin{equation}
{\rm Re}_{\rm M} \simeq
\frac{\sigma|\overline{\nabla}\times\vec{v}\times\vec{B}|}{|\nabla^2
\vec{B}|}\simeq v~L_{B}~\sigma.
\label{2.8}
\end{equation}
If $R\ll 1$ (for a given length scale $L_{B}$) the flux lines of the
magnetic field will diffuse through the plasma. If $R\gg 1$ the flux
lines of the magnetic field will be frozen into the plasma element.
From the magnetic diffusivity equation (\ref{mdiff}) it is possible to
derive the typical structure of the dynamo term by carefully averaging
over the velocity field according to the procedure outlined in  \cite{rev1,rev2}.
By assuming that the motion of the  fluid is random and that it has zero mean
velocity, it is possible to average over the ensemble of the possible
velocity fields.
In more physical terms this averaging procedure of Eq. (\ref{mdiff}) is
equivalent to averaging over scales and times exceeding the
characteristic correlation scale and time $\tau_{0}$ of the velocity
field. This procedure assumes that the correlation scale of the
magnetic field is much larger than the correlation scale of the
velocity field.
In this approximation the magnetic diffusivity equation can be written
as:
\begin{equation}
\frac{\partial\vec{B}}{\partial\eta} =
\alpha(\vec{\nabla}\times\vec{B}) +
\frac{1}{\sigma}\nabla^2\vec{B}, 
\label{2.9}
\end{equation}
($\alpha 
= -\frac{\tau_{0}}{3}\langle\vec{v}\cdot\vec{\nabla}
\times\vec{v}\rangle$ is the so-called dynamo term, which vanishes
in the absence of vorticity; in this equation $\vec{B}$ is
the magnetic field averaged 
over times larger than $\tau_{0}$, which is the typical correlation
time of the velocity field). The crucial
requirement for the  described averaging procedure is that the
turbulent velocity field has to be ``globally'' non-mirror-symmetric.
It is  interesting to point out \cite{rev2} that the
dynamo term in Eq. (\ref{2.9}) has a simple electrodynamical meaning,
namely, it can be interpreted as a mean ohmic current directed along
the magnetic field:
\begin{equation}
\vec{J} = - \alpha \vec{B}.
\label{2.10}
\end{equation}
This equation tells us that an ensemble of screw-like vortices with
zero mean helicity is able to generate loops in the magnetic flux
tubes in a plane orthogonal to the one of the original field.
This observation will be of some related interest for the physical
interpretation of the results we are going to present in the following
paragraph. We finally notice that if the velocity field {\em is}
parity-invariant (i.e. no vorticity for scales comparable with the
correlation length of the magnetic field), then the dynamics of the
infrared modes is decoupled from the velocity field since, over those
scales, $\alpha =0$. When the (averaged) dynamo term dominates in 
Eq. (\ref{2.9}), magnetic fields can be exponentially amplified. When 
the value of the magnetic field reaches the equipartition value 
(i.e. when the magnetic and kinetic energy of the plasma are comparable),
the dynamo ``saturates''. The precise way in which the dynamo effect
stops was a subject of active research some years ago \cite{rev1}. 
What happens is that close to equipartition, Eq. (\ref{2.9}) should be 
supplemented with non-linear terms whose effect is to stabilize
the amplification of the magnetic field to a constant value \cite{rev1}.

MHD equations can be studied in two different limits : the ideal (or 
superconducting) approximation where the conductivity is 
assumed to be very high and the real (or resistive) limit 
where the conductivity takes a finite value. 
In the ideal limit both the magnetic flux and the magnetic helicity 
are conserved. This means, formally \cite{mg1}, 
\begin{eqnarray}
\frac{d}{d\eta} \int_{\Sigma} \vec{B} \cdot d\vec{\Sigma}=-
\frac{1}{\sigma} \int_{\Sigma} \vec{\nabla} \times\vec{\nabla}
\times\vec{B}\cdot d\vec{\Sigma},
\label{flux}
\end{eqnarray}
where $\Sigma$ is an arbitrary closed surface which moves with the
plasma.

If we are in the in the inertial regime (i.e. $L>L_{\sigma}$ where 
$L_{\sigma } $ is the magnetic diffusivity length) we can
say that the expression appearing at the right hand side is
sub-leading
and the magnetic flux lines evolve glued to the plasma element.
 Today the magnetic diffusivity 
scale, estimated from MHD considerations, is of the 
order of the A. U.. This means that fields cohernt over 
distances smaller than the A. U. are dissipated. Conversely, if 
the coherence scale of the magnetic fields is larger 
than the A. U. the associated magnetic flux is conserved.

It is now worth mentioning  the magnetic helicity
\begin{equation}
{\cal H}_{M} = \int_{V} d^3 x \vec{A}~\cdot \vec{B},
\label{h1}
\end{equation}
where $\vec{A}$ is the vector potential \footnote{Notice that in conformally
flat FRW spaces the radiation gauge is conformally invariant. This property
is not shared by the Lorentz gauge condition \cite{ford}.}.
In Eq. (\ref{h1}) the vector potential appears and, therefore
it might seem that the expression is not gauge invariant. This is not
the case. In fact $\vec{A}\cdot\vec{B}$ is not gauge invariant but,
none the less, ${\cal H}_{M}$ is gauge-invariant since 
the integration volume is defined in such a way that the magnetic field
$\vec{B}$
is parallel to the surface which bounds $V$ and which we will call
$\partial V$. In $\vec{n}$ is the unit vector normal to $\partial V$
then $\vec{B}\cdot\vec{n}$ in $\partial V$. 

The magnetic gyrotropy
\begin{equation}
\vec{B}\cdot\vec{\nabla} \times\vec{B}
\end{equation}
it is a gauge invariant measure of the diffusion rate of ${\cal H}_{M}$
at finite conductivity. In fact \cite{mg1}
\begin{equation}
\frac{d}{d\eta} {\cal H}_{M} = - \frac{1}{\sigma} \int_{V} d^3 x
{}~\vec{B}\cdot\vec{\nabla} \times\vec{B}.
\label{h2}
\end{equation}
The magnetic gyrotropy is a useful quantity in order to distinguish 
different mechanisms for the magnetic field generation. Some 
mechanisms are able to produce magnetic fields whose 
flux lines have a topologically non-trivial structure (i.e. 
$\langle \vec{B} \cdot \vec{\nabla} \times \vec{B} \rangle \neq 0$).
This observation will be used in the following Sections.

The discovery of large scale magnetic fields in the intra-cluster medium 
implies some interesting problems for the mechanisms of generation 
of large scale magnetic fields. Consider, first, magnetic fields in galaxies. 
Usually the picture 
for the formation of galactic magnetic fields is related to the 
possibility of implementing the dynamo mechanism.  
By comparing 
the rotation period with the age of the galaxy (for a Universe with 
$\Omega_{\Lambda} \sim 
0.7$, $h \sim 0.65$ and $\Omega_{\rm m} \sim 0.3$) the number of rotations
performed by the galaxy since its origin is approximately $30$. 
During these $30$ rotations the dynamo term of Eq. (\ref{mdiff}) 
 dominates against the magnetic diffusivity term since parity 
is globally broken over the physical size of the galaxy. As a 
consequence an instability develops. This instability can be used
in order to drive the magnetic field from some small initial condition
up to its observed value. Most of the work in the context of the dynamo 
theory focuses on reproducing the correct features of the 
magnetic field of our galaxy.
For instance one could ask the dynamo codes to reproduce the 
specific ratio between the 
 poloidal and toroidal amplitudes of the magnetic field of the Milky Way. 

The achievable amplification produced by the 
dynamo instability can be at most of $10^{13}$, i.e. $e^{30}$. Thus, if 
the present value of the galactic magnetic field is $10^{-6}$ Gauss, its value 
right after the gravitational collapse of the protogalaxy might have 
been as small as $10^{-19}$ Gauss over a typical scale of $30$--$100$ kpc.

There is a simple way to relate the value of the magnetic fields 
right after gravitational collapse to the value of the magnetic field 
right before gravitational collapse. Since the gravitational collapse 
occurs at high conductivity the magnetic flux and the magnetic helicity
are both conserved. Right before the formation of the galaxy a patch 
of matter of roughly $1$ Mpc collapses by gravitational 
instability. Right before the collapse the mean energy density  
of the patch, stored in matter, 
 is of the order of the critical density of the Universe. 
Right after collapse the mean matter density of the protogalaxy
is, approximately, six orders of magnitude larger than the critical density.

Since the physical size of the patch decreases from $1$ Mpc to 
$30$ kpc the magnetic field increases, because of flux conservation, 
of a factor $(\rho_{\rm a}/\rho_{\rm b})^{2/3} \sim 10^{4}$ 
where $\rho_{\rm a}$ and $\rho_{\rm b}$ are, respectively the energy densities 
right after and right before gravitational collapse. Henceforth, the 
correct initial condition in order to turn on the dynamo instability
is $B \sim 10^{-23}$ Gauss over a scale of $1$ Mpc, right before 
gravitational collapse. 

Since the flux is conserved the ratio between the magnetic energy 
density, $\rho_{\rm B}(L,\eta)$ 
 and the energy density sitting in radiation, $\rho_{\gamma}(\eta)$
 is almost constant and therefore, in terms of this quantity (which is only scale 
dependent but not time dependent), the dynamo requirement can be rephrased as
\begin{equation}
r_{\rm B}(L) = \frac{\rho_{\rm B}(L,\eta)}{\rho_{\gamma}(\eta)} 
\geq 10^{-34},\,\,\,\, L\sim 1\,{\rm Mpc}.
\label{dyn}
\end{equation}
If the dynamo is not invoked but the galactic magnetic field directly 
generated through some mechanism the correct value to impose 
at the onset of gravitational collapse is much larger.
The possible applications of dynamo mechanism to  clusters is still
under debate and it seems more problematic \cite{cl5,cl6,dc}.  
The typical scale of the gravitational collapse of a cluster 
is larger (roughly by one order of magnitude) than the scale of gravitational
collapse of the protogalaxy. Furthermore, the mean mass density 
within the Abell radius ( $\simeq 1.5 h^{-1} $ Mpc) is roughly 
$10^{3}$ larger than the critical density. Consequently, clusters 
rotate less than galaxies since their origin and the value of 
$r_{\rm B}(L)$ has to be larger than in the case of galaxies. 
Since the details of the dynamo mechanism applied to clusters are 
not clear, at present, it will be required that $r_{\rm B}(L) \gg 10^{-34}$
[for instance $r_{\rm B} (L) \simeq 10^{-12}$]. 

\renewcommand{\theequation}{3.\arabic{equation}}
\setcounter{equation}{0}
\section{Two classes of mechanisms}

In the context of the  ideas illustrated in the previous section,
large scale galactic magnetic fields are assumed to 
be the result of the amplification of a primeval seed. 
It was Harrison 
\cite{second} who suggested that these seeds might have something 
to do with cosmology in the same way as 
he  suggested that the primordial spectrum of gravitational potential 
fluctuations (i.e. the Harrison-Zeldovich spectrum) 
might be produced in some primordial phase of the evolution of the Universe.
Since then, several mechanisms have been invoked 
in order to explain the origin of these seeds \cite{s1,s3} and few of them 
are compatible with inflationary evolution. It is not my purpose 
to review here all the different mechanisms which have been proposed and 
good reviews exist already \cite{en,rub,dol} (see also \cite{pro}). 
A very incomplete 
selection of references is, however, reported \cite{s1,s2,s3}.  Here I would 
like to summarize the situation in light of the recent theoretical 
and experimental developments. 
Consider first of all the case of the galaxy. 
The mechanisms for magnetic field generation can be divided,
broadly speaking, into two categories: astrophysical \cite{rev3,kul} and 
cosmological. The cosmological mechanisms can be divided, in their turn,
into {\em causal} mechanisms (where the magnetic seeds are produced at a given time inside the 
horizon) and {\em inflationary} mechanisms where correlations in the magnetic field 
are produced outside the horizon. Astrophysical mechanisms 
have always to explain the initial conditions of Eq. (\ref{mdiff}). This is 
because the MHD are linear in the magnetic fields. It is 
questionable if  purely astrophysical considerations 
can set a natural initial condition for the dynamo amplification.

\subsection{Turbulence?}
Both classes of mechanisms have problems. Causal mechanisms usually fail 
in reproducing the correct correlation scale of the field whereas 
inflationary mechanisms have problems in reproducing the correct amplitude
required in order to turn on successfully the dynamo action.
In the context of causal mechanisms there are interesting 
proposals in order to enlarge the correlation scale. These 
proposals have to do with the possible occurrence of turbulence 
in the early Universe.  In order to discuss  
the turbulent features of a magnetized plasma 
 the kinetic Reynolds number
\begin{equation}
{\rm Re} =\frac{v~L_{v}}{\nu} 
\label{re}
\end{equation}
should be employed together with the magnetic Reynolds number 
defined in Eq. (\ref{2.8}) 
from the relative balance of the dynamo and diffusivity
terms. In Eq. (\ref{re}) $\nu$ is the thermal diffusivity coefficient, $v$ the bulk velocity
of the plasma and $L_{v}$ the typical correlation scale 
of the velocity field.
The ratio of the magnetic Reynolds number 
to the kinetic Reynolds number is the Prandtl number \cite{bis}
\begin{equation}
{\rm Pr}_{\rm M} = \frac{{\rm Re}_{\rm M}}{{\rm Re}} = \nu \sigma 
\biggl( \frac{L_{B}}{L_{v}} \biggr).
\label{pr}
\end{equation}
Notice that in Eqs. (\ref{2.8}) and (\ref{re}) the correlations scales 
of the magnetic field and of the fluid motion are, in principle, 
different. Indeed, as discussed in connection with 
the coarse-grained dynamo equation (\ref{2.9}), the correlation scale 
of the fluid motions should be, initially, shorter than the 
magnetic correlation scale. Even assuming that $L_{B} \simeq L_{v}$  
the Prandtl number is typically much larger than one. 
Consider, for instance, the case of the electroweak epoch 
\cite{turb1,turb2,turb3,mg3,mg4,turb4}. At this epoch 
taking $H_{\rm ew}^{-1} \sim 3 {\rm cm}$ we get that 
${\rm Pr}_{\rm M}\sim \nu\sigma \sim 10^{6} $ where the bulk velocity 
of the plasma is of the order of the bubble wall velocity 
at the epoch of the phase transition. 

This means that the early universe is both kinetically and magnetically 
turbulent. The features of magnetic and kinetic 
turbulence are different. This aspect reflects in 
a spectrum of fluctuations is different from the usual 
Kolmogorov spectrum \cite{bis}. If the Universe is both 
magnetically and kinetically turbulent it has been speculated 
that an inverse cascade mechanism can occur 
\cite{turb1,turb2,turb3,turb4}. This idea was originally 
put forward in the context of MHD simulations \cite{bis}. 
The inverse cascade would imply a growth in the correlation scale 
of the magnetic inhomogeneities and it has been shown to 
occur numerically in the approximation of unitary 
Prandtl number \cite{bis}. Specific cascade models 
have been also studied \cite{turb1,turb2,turb3,turb5a}. 
A particularly important r\^ole is played, in this context, 
by the initial spectrum of magnetic fields (the so-called 
injection spectrum \cite{turb3}) and by the topological 
properties of the magnetic flux lines. In particular, if the 
system has non vanishing magnetic helicity and magnetic gyrotropy
it was suggested that the inverse cascade can occur more efficiently 
\cite{turb5,turb6}. Recently simulations have discussed 
the possibility of inverse cascade in realistic MHD models \cite{turb6}. 
More analytic discussions based on renormalization group 
approach applied to turbulent MHD seem to be not totally consistent with
the occurrence of inverse cascade at large scales \cite{bere}.

\subsection{Magnetic fields from vacuum fluctuations}

Large scale magnetic fluctuations can be generated 
during the early history of the Universe and can go 
outside the horizon with a mechanism similar 
to the one required in order to produce fluctuations 
in the gravitational potential. In this case the correlation 
scale of the magnetic inhomogeneities can be large. However, the 
typical amplitudes obtainable in this class of models may be too small.
The key property allowing the amplification of the fluctuations 
of the scalar and tensor modes of the geometry is the fact 
that the corresponding equations of motion are not invariant 
under Weyl rescaling of a (conformally flat) metric of FRW type.
In this sense the evolution equations of relic gravitons 
and of the scalar modes of the geometry are not conformally 
invariant. The evolution equations of gauge fields do not share this 
property. However, if gauge couplings are dynamical, there 
is a natural way of breaking conformal invariance in the
evolution equation of gauge fields. 
Interesting examples in this direction are models containing 
extra-dimensions and scalar-tensor theories of gravity 
where the gauge coupling is, effectively, a scalar 
degree of freedom evolving in a given geometry.

\renewcommand{\theequation}{4.\arabic{equation}}
\setcounter{equation}{0}
\section{Dynamical extra-dimensions} 
The remarkable similarity of the 
abundances of light elements in different 
galaxies leads to postulate that the 
Universe had to be dominated by radiation 
at the moment when the light elements were formed, 
namely for temperatures of approximately $0.1 $ MeV 
\cite{sar1,sar2}. 
Prior to the moment of nucleosynthesis 
even indirect informations concerning the thermodynamical 
state of our Universe are lacking even if our knowledge 
of particle physics could give us important hints concerning 
the dynamics of the electroweak phase transition \cite{mis}.
 
The success  of big-bang nucleosynthesis (BBN) 
sets limits on  alternative cosmological scenarios. 
Departures from homogeneity \cite{hom} and isotropy \cite{iso} 
of the background 
geometry can be successfully constrained. 
Bounds on  the presence of matter--antimatter domains of various sizes
can be derived \cite{anti1,anti2,anti3}. 
BBN can also set limits on the 
dynamical evolution of internal dimensions \cite{int1,int2}.
Internal dimensions are an essential ingredient 
of  theories attempting the unification of gravitational and 
gauge interactions in a higher dimensional background like 
Kaluza-Klein theories \cite{kk}  and superstring theories \cite{ss}. 
 
Defining, respectively, $b_{BBN}$ 
and $b_0$ as the size of  the internal dimensions at the BBN 
time and at the present epoch, the maximal variation 
allowed to the internal scale factor from the BBN time 
can be expressed as $b_{BBN}/b_0 \sim 1 
+ \epsilon$ where $ |\epsilon | < 10^{-2} $ \cite{int1,int2}. 
The bounds on the  variation 
of the internal dimensions during the matter dominated epoch
are even stronger. Denoting with an over-dot the derivation with 
respect to the cosmic time coordinate, we have that 
$ |\dot{b}/b| < 10^{-9} H_0$ where 
$H_0$ is the present value of the Hubble parameter \cite{int1}.
The fact that the time evolution of internal dimensions
is so tightly constrained for temperatures lower of $1$ MeV
does not forbid that they could have been dynamical 
prior to that epoch. Moreover, recent observational 
evidence \cite{we1,we2,we3} seem to imply that 
the fine structure constant can be changing even today.

Suppose that prior to BBN internal dimensions 
were evolving in time  and assume, for sake of simplicity, that 
after BBN the internal dimensions have been frozen to their present 
(constant) value. 
Consider a homogeneous and anisotropic manifold 
whose line element can be written as 
\begin{eqnarray}
ds^2 = G_{\mu\nu} dx^{\mu} dx^{\nu} = 
a^2(\eta) [ d\eta^2 - \gamma_{i j} d x^{i} dx^{j}] - b^2(\eta) \gamma_{a b} 
dy^a d y^b,
\nonumber \\
\mu,\nu = 0,..., D-1=d+n , ~~~~~~ i, j=1,..., d , ~~~~~~ 
a,b = d+1,..., d+n.
\label{metric}
\end{eqnarray}
[$\eta$ is the
conformal time coordinate related, as usual to the cosmic time $t=\int a(\eta)
d\eta$ ; $\gamma_{ij}(x)$, $\gamma_{ab}(y)$ are the metric
tensors of two maximally symmetric Euclidean 
manifolds parameterized,
respectively, by the ``internal" and the ``external" coordinates $\{x^i\}$ and
$\{y^a\}$]. 
The metric of Eq. (\ref{metric})
 describes the situation in which the $d$ external dimensions 
(evolving
with scale factor $a(\eta)$) and  the $n$ internal ones (evolving with scale
factor $b(\eta)$) are dynamically decoupled from each other \cite{gio1}. 
The results of the present investigation, however,
can be easily generalized to the case of $n$ different scale factors in the 
internal manifold.

Consider now  a pure electromagnetic fluctuation decoupled 
from the sources, representing an electromagnetic wave propagating 
in the $d$-dimensional external space such that $A_{\mu} \equiv A_{\mu}(\vec{x}, 
\eta)$, 
$A_{a} =0$. In the metric given in Eq. (\ref{metric}) the evolution 
equation of the gauge field fluctuations can be written as 
\begin{equation}
\frac{1}{\sqrt{-G}} \partial_{\mu}\biggl( \sqrt{-G} G^{\alpha\mu} 
G^{\beta\nu} F_{\alpha\beta} \biggr) =0,
\end{equation}
where $F_{\alpha\beta} = \nabla_{[\alpha}A_{\beta]}$ 
is the gauge field strength and 
$G$ is the determinant of the $D$ dimensional metric. Notice that 
if $n=0$ the space-time is isotropic and, therefore, the Maxwell's 
equations can be reduced (by trivial rescaling) to the flat space equations. 
If $n \neq 0$ we have that 
the evolution equation of the electromagnetic fluctuations propagating in the 
external $d$-dimensional manifold will receive a contribution from the internal
 dimensions which cannot be rescaled away. 

In the radiation gauge ($A_0 =0$ and $\nabla_{i} A^{i} =0$) 
the evolution  the vector potentials can be written as 
\begin{equation}
A_{i}'' + n {\cal F} A_{i}' - \vec{\nabla}^2 A_{i} =0, \,\,\,\,\,
 {\cal F} = \frac{b'}{b}.
\label{vec1}
\end{equation}
The vector potentials $A_{i}$ are already rescaled with respect 
to the (conformally flat) $d+1$ dimensional metric.  
In terms of the canonical normal modes of oscillations ${\cal A}_{i} = b^{n/2}
 A_{i}$ 
the previous equation can be written in a simpler form, namely 
\begin{equation}
{\cal A}_{i}''  - V(\eta) {\cal A}_{i} -\vec{\nabla}^2 {\cal A}_{i}  =0,\,\,\,
\,\, 
V(\eta) 
= \frac{n^2}{4} {\cal F}^2 + \frac{n}{2}{\cal F}'.
\label{eq1}
\end{equation}

From this set of equations the induced large scale magnetic fields 
can be computed in various models for the evolution 
of the internal manifold \cite{mg}.

\renewcommand{\theequation}{5.\arabic{equation}}
\setcounter{equation}{0}
\section{Dynamical gauge couplings in four-dimensions}

The evolution of the (Abelian) 
gauge coupling during an inflationary phase of de 
Sitter type drives the growth of the two-point function
of the magnetic inhomogeneities \cite{dinng,dinng2}.
The idea that a theory with local gauge-invariance could lead to 
a consistent variation of the Abelian coupling was explored by Dirac
and, subsequently, by Teller and Bekenstein \cite{dir1,dir2,dir3,dir4}. 
This physical possibility is also realized in superstring-inspired 
cosmological scenarios where the gauge coupling is related to the 
expectation value of the dilaton field \cite{pbb1}. In the following we propose 
a model for the evolution of the 
gauge coupling in a standard cosmological scenario where 
the inflationary phase is followed by a radiation and a matter dominated epoch. 
  
Suppose that a minimally coupled (massive) scalar field $\phi$ evolves in a 
conformally flat metric of FRW type (\ref{metric1}). 
The field $\phi$ {\em is not the inflaton} but it evolves 
during different cosmological epochs parametrized by a 
different form of $a(\eta)$. Typically the Universe 
evolves from an inflationary phase of de Sitter (or quasi-de Sitter)
 type towards
a radiation dominated phase which is finally replaced
by a matter dominated epoch.

The evolution equation of $\phi$ in the background given by Eq. (\ref{metric}) 
can be written as 
\begin{equation}
\phi'' + 2 {\cal H} \phi' + m^2 a^2 \phi=0,\,\,\,{\cal H} = \frac{a'}{a},
\label{ph}
\end{equation}
where 
${\cal H}$ is the Hubble factor in conformal time related to 
the Hubble parameter in cosmic time $H = \dot{a}/a$ as $ H a ={\cal H}$
(the dot denotes derivative with respect to cosmic time).

If $\phi$ evolves during an inflationary phase 
of de Sitter type the scale factor will be  
\begin{equation}
a(\eta) = \bigl( -\frac{\eta}{\eta_1})^{-1}, \,\,\,\eta < - \eta_1,
\label{inf}
\end{equation}
where $-\eta_1$ marks the end of the inflationary phase. 
If $m^2 a^2 \ll {\cal H}$ (i.e. $m \ll H$) during inflation, according 
to Eq. (\ref{ph}),  $\phi$ relaxes as 
$\phi \sim (-\eta/\eta_1)^3$ for $\eta < - \eta_1$.

Suppose, as an example, 
 that $\phi$ is coupled to an (Abelian) gauge field  
\begin{equation}
S \sim \int d^4 x \sqrt{-G} \phi^2 F_{\mu\nu}F^{\mu\nu}.
\label{ac}
\end{equation}
The normal modes of the hypermagnetic field fluctuations 
$B_{i}(\vec{x}, \eta)$ are 
\begin{equation}
b_{i}(\vec{x},\eta) = \phi(\eta) B_{i}(\vec{x},\eta),
\end{equation}
 and their 
correlation function during the de Sitter phase can then be written as
\begin{equation}
{\cal G}_{ij} (\vec{r},\eta) = \int \frac{d^3 k}{(2\pi)^3} P_{i j}(k)
 b(k,\eta) b^{\ast}(k\eta) 
e^{i \vec{k}\cdot \vec{r}},
\label{two}
\end{equation}
where 
\begin{equation}
P_{i j} = \biggl( \delta_{i j} - \frac{k_{i} k_{j}}{k^2} \biggr).
\end{equation}
The normal modes  $b(k,\eta)$ will evolve as 
\begin{equation}
b'' +\bigl[k^2 - \frac{\phi''}{\phi}\bigr] b =0.
\end{equation}
Using now the fact that $\phi \sim \eta^3$ the correlation function, 
during the de Sitter phase grows as
\begin{equation}
{\cal G}_{i j} (\vec{r},\eta) \sim (-\eta)^{-4},
\end{equation}
for $\eta \rightarrow 0^{-}$ (i.e. $t\rightarrow\infty$). 
Thus, gauge field fluctuations grow during the Sitter stage. Furthermore,
from Eq. (\ref{ac}) the magnetic energy density $\rho_{\rm B}(r,\eta)$
[related to the 
trace of ${\cal G}_{ij}(\vec{r},\eta)$]  also
increases for $\eta\rightarrow 0^{-}$. Consequently, since
the magnetic energy density can be amplified during a de Sitter-like 
stage of expansion, large scale gauge fluctuations 
pushed outside of the horizon can generate the galactic magnetic field. 

\subsection{Evolution of the gauge coupling}

The only gauge coupling free to evolve, in the present discussion, 
 is the one associated 
with the hypercharge field leading, after symmetry breaking, to the time 
variation of the electron charge. In 
 a relativistic plasma the conductivity goes,  
approximately, as $T/\alpha_{\rm em}$ where $\alpha_{\rm em}$ 
is the fine structure constant . If $\alpha_{\rm em}$ depends on time 
also the well known magnetohydrodynamical equations (MHD). 
will have to be generalized,  leading, ultimately, 
to different mechanism for the relaxation of the 
magnetic fields. 

If the evolution of the Abelian coupling is parametrized through the 
minimally coupled scalar field $\phi$, the possible 
constraints pertaining to the evolution of $\phi$ are translated 
into constraints on the evolution of the gauge coupling.
Massless scalars cannot exist in the Universe: they lead 
to long range forces whose effect should appear in 
sub-millimiter tests of Newton's law. 
Consequently, the scalar 
mass should be, at least, larger than $10^{-4}$ eV otherwise 
it would be already excluded \cite{mm1,mm2}. 
Massive scalars are severely 
constrained from cosmology \cite{r,l,l2}. When 
the scalar mass is comparable with the Hubble 
rate (i.e. $m \sim H$) the 
field starts oscillating coherently 
with Planckian amplitude and $\phi$
decays too late big-bang nucleosynthesis (BBN) 
can be spoiled.

Initial conditions for the evolution 
of $\phi$ are given during a de Sitter 
stage of expansion. Thus, the homogeneous evolution of 
$\phi$ can be written as 
\begin{equation}
\phi_{\rm i}(\eta) \sim \phi_1 - \phi_2 \bigl( \frac{\eta}{\eta_1}
 \bigr)^3,\,\,\,\eta < -\eta_1,
\end{equation}
where $\phi_1$ is the asymptotic value of $\phi$ which may or may 
not coincide with the minimum of $V(\phi)$; $\phi_2$ is 
also an integration constant. Without fine-tuning 
$\phi_1$ and $\phi_2$ both coincide with $M_{P}$.
During the inflationary phase 
$m\ll {\cal H}_1/a_1 = H_1$ where, $H_1 < 10^{-6} \,M_{P}$ is the 
curvature scale at the end of inflation.

After $\eta_1$ the Universe enters a phase of radiation
dominated evolution  (possibly preceded by a 
reheating phase) where the curvature scale 
decreases. When $H_{\rm m} \sim m$  
the scalar field starts oscillating coherently 
with  amplitude $\phi_1$. 

During reheating the scale factor evolves 
as $a(\eta) \sim \eta^{\alpha}$ so that, in this 
phase, $\phi$ relaxes as 
\begin{equation}
\phi_{\rm rh} 
\sim \phi_1 + \bigl(\frac{\eta}{\eta_1})^{ 1 - 2\alpha},\,\,\eta_1 < 
\eta <\eta_r,
\label{rh}
\end{equation}
where $\eta_{\rm r}$ marks the beginning of the radiation 
dominated phase occurring at a scale $H_{\rm r}>m$. 
 In the case of matter-dominated equation of 
state during reheating $\alpha \sim 2 $.

For $\eta>\eta_r$, the evolution of the field 
$\phi$ can be exactly solved 
(in cosmic time) in terms of Bessel functions
\begin{equation}
\phi(t) \sim a^{-3/2}(t) 
\sqrt{m t} \bigl[ A Y_{1/4}(m t) + B J_{1/4}(m t)\bigr],
\label{ra}
\end{equation}
where $A$ and $B$ are two integration constants.
From Eq. (\ref{ra}), $\phi \sim {\rm constant} +\eta^{-1} $
for $H> m$, and it oscillates for $H < m$. 
When $H <m$ the coherent oscillations of $\phi$ start and 
their energy density  decreases as
$a^{-3}$.
The curvature scale $H_{\rm c}$ marks the time 
at which the energy density stored in the 
 coherent oscillations equal 
the energy density of the radiation background,  namely
\begin{equation}
H_{\rm r}^2 M_{P}^2 \bigl(\frac{a_{\rm r}}{a_{\rm c}}\bigr)^4 \sim 
m^2 \phi_1^2 \bigl(\frac{a_{\rm m}}{a_{\rm c}}\bigr)^3,
\label{eq}
\end{equation}
where $\eta_{\rm m}$  
correspond to the times at which $H\sim m$. From Eq. (\ref{eq})
\begin{equation}
H_{\rm c} \sim \xi\, \varphi^4\,M_{P}.
\end{equation}
where $\varphi = \phi_1/M_{P}$ and $\xi = m/M_{P}$.
The phase of dominance of coherent oscillation ends with
the decay of $\phi$ at a scale dictated by the strength 
of gravitational interactions and by the mass $m$, namely
\begin{equation}
H_{\phi} \sim \xi^3 \,M_{P}.
\end{equation}
In order not to spoil the light elements abundances 
we have to require that $H_{\phi} > H_{\rm ns} $
implying that $ m > 10$ TeV.  

In order not wash-out the baryon asymmetry produced
at the electroweak time by overproduction of entropy \cite{ew1,ew2}
$H_{\phi} > H_{\rm ew}$ may be imposed. Since 
$H_{\rm ew} \sim \sqrt{N_{\rm eff}} T^2_{\rm ew} / M_{P}$ 
[where $N_{\rm eff}=106.75$ is the effective number of (spin)
 degrees of freedom at $T_{\rm ew} \sim 100$ GeV] 
 we obtain $ m > 10^{5} $ TeV.
Notice, incidentally, that the time variation of the gauge 
couplings during the electroweak epoch 
 has not been analyzed and it may be relevant 
in order to produce inhomogeneities at the onset of BBN.

The inhomogeneous modes of 
$\phi$ should also be taken into account since we have 
to check that  
further constraints are not introduced. In order 
to find how many quanta of the field $\phi$ are produced 
by passing from the inflationary phase to a radiation 
dominated phase let us look at the sudden approximation for 
the transition of $a(\eta)$ \cite{ms}. 
Consider the first order fluctuations of the field $\vf$
\begin{equation}
\vf(\vec{x},\eta) = \vf(\eta) + \delta \vf(\vec{x},\eta),
\end{equation}
whose evolution equation is, in Fourier space,
\begin{equation}
\psi'' + 2 {\cal H} \psi' + [ k^2 + m^2 a^2] \psi =0,
\label{fluc}
\end{equation}
where $\psi(k,\eta)$ is the Fourier component 
of $\delta\vf(\vec{x}, \eta)$. 

In the limit $ k\eta_1 \ll 1$ the mean number of 
quanta created by parametric amplification of vacuum 
fluctuations \cite{ms} is
\begin{equation}
\overline{n}(k) \simeq |c_{-}(k)|^2= q |k \eta_1|^{-2 \lambda} 
\biggl(\frac{m}{H_1}\biggr)^{-1/2} 
\end{equation}
where $q$ is a numerical coefficient of the 
order of $10^{-2}$. 
the energy density of the created (massive) quanta 
can be estimated from 
\begin{equation}
d\rho_{\psi} = \frac{d^3 \omega}{(2\pi)^3} m \overline{n}(k),
\end{equation}
where $ \omega = k/a$  is the physical momentum. 
In the case of a de Sitter phase ($\lambda = 3/2$) the typical 
energy density of the produced fluctuations 
is 
\begin{equation}
\rho_{\psi}(\eta) \simeq 
q \,m\, H_1^3\, 
\biggl( \frac{m}{H_1}\biggr)^{-1/2} \biggl(\frac{a_1}{a}\biggr)^3
\end{equation}
Also the massive fluctuations may become dominant and we have to make 
sure that they become dominant after $\vf$ already decayed. Define 
$H_{\ast}$ as the scale at which the massive fluctuations become 
dominant with  respect to the radiation background.
The scale $H_{\ast}$ can be determined 
by requiring that $\rho_{\psi}(\eta_{\ast}) \simeq \rho_{\gamma} (\eta_{\ast})$
implying that 
\begin{equation}
m \,H_1^3 \biggl(\frac{m}{H_1}\biggr)^{-1/2} 
\biggl(\frac{a_1}{a_{\ast}}\biggr)^3 
\simeq H_1^2 \, M_{P}^2 \biggl(\frac{a_1}{a_{\ast}}\biggr)^4,
\end{equation}
which translates into
\begin{equation}
H_{\ast}  \simeq q \, \xi\,\epsilon^4\, M_{P},
\end{equation}
where $\epsilon= H_1/M_{P}$.
In order to make sure that the non-relativistic modes 
will become dominant after $\vf$ already decayed 
we have to impose that $H_{\ast} < H_{\vf}$ which means that
$m > 10^{2}$ TeV for $H_1\sim 10^{-6}\, M_{P}$ and which is less 
restrictive than the other constraints previously derived in this paper.

\subsection{Magnetogenesis}
The full action describing the problem of the evolution 
of the gauge coupling  in this simplified scenario is 
\begin{equation}
S= \int d^4 x \,\,\sqrt{-G} \biggl[ \frac{1}{2} G^{\mu\nu} \partial_{\mu} 
\vf \partial_{\nu} \vf 
- V(\vf) - \frac{1}{4} f(\phi) F_{\mu\nu} F^{\mu\nu} 
\biggr].
\label{action1}
\end{equation}
Using Eq. (\ref{metric}) the equations of motion become
\begin{eqnarray}
&& \vf'' +  2 {\cal H} \vf ' +a^2 \frac{\partial V}{\partial \phi} 
= -\frac{1}{2 a^2}
\frac{\partial f}{\partial \phi} \biggl[ \vec{B}^2 - \vec{E}^2 \biggr] 
\label{s1}\\
&& \frac{\partial{\vec{B}}}{\partial\eta} = -\vec{\nabla}
\times \vec{{E}} 
,~~~~~{\vec{\nabla}}\cdot {\vec{E}}=0,
\label{s1b}\\
&&\frac{\partial}{\partial\eta}\biggl[f(\phi) \vec{E}\biggr] +
 \vec{J} =
f(\phi) {\vec{\nabla}}\times \vec{B}, 
\label{s2}\\
&& {\vec{\nabla}}\cdot{\vec{B}}=0,~~~~~~
\vec{J}=\sigma ({\vec{E}} + \vec{v}\times{\vec{B}})
\label{s4}
\end{eqnarray}
(${\vec{B}}=a^2 \vec{{\cal B }}$, $\vec{E}=a^2
\vec{{\cal E}}$; $\vec{J}=a^3 \vec{j}$; $\sigma=
\sigma_{c} a$;  $\vec{{\cal
B}}$, $\vec{{\cal E}}$,
$\vec{j}$, $\sigma_{c}$ are  the flat-space quantities whereas
$\vec{B}$, $\vec{E}$, $\vec{J}$, $\sigma$ are the
curved-space ones; $\vec{v}$ is the bulk velocity of the plasma). 

In Eqs. (\ref{s1})--(\ref{s4}) the effect of the conductivity 
has been included. The current density [present in Eq. (\ref{s1}) with a term 
$(\partial j_{\alpha}/\partial\phi) A^{\alpha}$] has been eliminated by the 
using Maxwell's equations. During 
the inflationary phase, for $\eta < -\eta_1$,
the role of the conductivity 
shall be neglected. In this case 
the evolution equation for 
the canonical normal modes of the magnetic field
can be derived from the curl of Eq. (\ref{s2}) 
with the use of Eq. (\ref{s1b}): 
\begin{equation}
\vec{b}'' - \nabla^2 \vec{b} - \bigl[ \frac{1}{2}\frac{f''}{f} - 
\frac{1}{4} \bigl(\frac{f'}{f}\bigr)^2\bigr] \vec{b} =0,
\label{b}
\end{equation}
where $\vec{b} = \sqrt{f} \vec{B}$.
For $\eta > - \eta_1$ the effect of the conductivity
is essential. Therefore, the correct 
equations obeyed by the magnetic field will be 
the generalization of the MHD equations whose 
derivation will be now outlined.

MHD equations represent an effective description 
of the plasma dynamics for large length scales 
(compared to the Debye radius) and 
short frequencies compared to the 
plasma frequency. MHD can be derived 
from the kinetic (Vlasov-Landau) equations 
and the MHD spectrum 
indeed reproduces the plasma spectrum 
up to the Alvf\'en frequency. 
MHD 
can be also derived  by neglecting the 
displacement currents in Eq. (\ref{s2}):
\begin{equation}
f \vec{\nabla}\times \vec{B} = \vec{J} + f' \vec{E}.
\end{equation}
By now using the Ohm law together 
with the Bianchi identity
we get to 
\begin{equation}
\bigl( 1 +  \frac{f'}{sf}\bigr) \vec{B}' = 
\vec{\nabla}\times(\vec{v} \times \vec{B}) + 
\frac{1}{s} \nabla^2 \vec{B}
\label{mh}
\end{equation}
which is the generalization of MHD 
equations to the case of evolving gauge coupling.
The quantity $s = \sigma /f$ is {\em constant}. 
The reason for this 
statement is the following. The rescaled conductivity,
\begin{equation}
\sigma = \sigma_c a \equiv \frac{T}{\alpha_{\rm em}},
\end{equation}
where $\alpha_{\rm em} \sim f^{-1} $. Therefore $\sigma/f =s$ 
with these rescalings, is constant. Taking 
now the Fourier transform of the fields appearing 
in Eq. (\ref{mh}) the solution, for the 
Fourier modes, will be 
\begin{equation}
B_{i}(k,\eta) = B_{i}(k,\eta_1) e^{- \int \frac{k^2 f}{ s f + f'} d\eta}.
\end{equation}

Consider now, as an example,  
\begin{equation}
f(\phi) = \bigl( \frac{\phi - \phi_1}{M_{P}}\bigr)^2.
\label{f}
\end{equation}
For $\eta < - \eta_1$ the solution of the evolution 
equation of the magnetic fluctuations
is, from Eq. (\ref{b}),
\begin{equation}
b(k, \eta) = N \sqrt{k \eta} H^{(2)}_{\nu}(k\eta),\,\,\, 
N= \frac{\sqrt{k\pi}}{2} e^{- 
i\frac{\pi}{4}(1 + 2\nu)},
\label{norm}
\end{equation}
where $N$ has been chosen in such a way that $b(k,\eta)\rightarrow 
\sqrt{k/2} e^{- i k
\eta}$ for $\eta\rightarrow -\infty$. Using Eq. (\ref{f}) 
 $\nu = 5/2$.

For $\eta_1 < \eta < \eta_{r}$ the Universe is reheating. 
During this phase the conductivity is not yet dominant
and the fastest growing solution outside the horizon
is given, in the case of Eq. (\ref{f}), by 
\begin{equation}
b(\eta) \sim \sqrt{f} \int_{\eta_1}^{\eta_{\rm r}} \frac{d\eta}{f}, 
\end{equation}
where we assumed, for concreteness, that $\alpha =2 $ in Eq. (\ref{rh}).
For $\eta > \eta_{\rm r}$ Eqs. (\ref{mh}) should be used.

The typical present frequency corresponding to the end of the inflationary 
phase is given, at the present time $\eta_0$, by 
\begin{equation}
\omega_{1}(\eta_0) \sim 10^{-4}\,\,T_{\rm dec}\,\,  
\epsilon^{\frac{1}{\alpha + 1}}\, 
\zeta^{\frac{\alpha -1}{2(\alpha + 1)}} \xi^{1/3}\,
\varphi^{- 2/3},
\label{om1}
\end{equation}
Notice that $10^{-4}\,T_{\rm dec}=100$ GHz where $T_{\rm dec}$ 
is the decoupling 
temperature. In Eq. (\ref{om1}), $\zeta = H_{\rm r}/M_{P}$ 
where $H_{\rm r}$ is the typical curvature scale associated
with $\eta_{\rm r}$ [see e.g. Eqs. (\ref{ra})--(\ref{eq})].
In the case of instantaneous reheating $H_{1} \simeq H_{\rm r}$.
The typical (present) frequency corresponding to the 
onset of the radiation dominated phase is given by
\begin{equation}
\frac{\omega_{\rm r}(\eta_0)}{\omega_1(\eta_0)} \sim 
\biggl(\frac{\zeta}{\epsilon}\biggr)^{ \frac{1}{\alpha + 1}}.
\end{equation}

\begin{figure}[htb]
    \centering
    \includegraphics[height=2.5in]{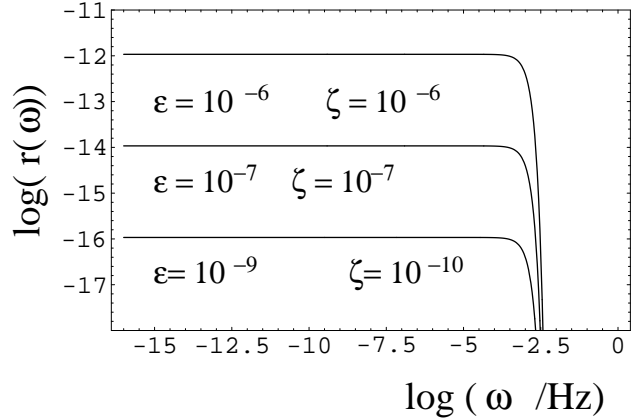}
    \caption{Different spectra are reported for various choices 
of the parameters. The value of $\xi$ has been fixed to $10^{-12}$ 
in order to have $m\geq 10^{5} $TeV, as required by 
electroweak baryogenesis considerations.}
    \label{F2}
\end{figure}

For $\eta > \eta_{\rm r}$ the conductivity 
dominates the evolution and using Eq. (\ref{mh}) we can estimate
the trace of the two-point function (\ref{two}) 
\begin{equation}
\rho_{B}(r,\eta) = \int \rho_{B}(k, \eta) \frac{ \sin{kr}}{k r} \frac{ dk}{k},
\end{equation}
with
\begin{equation}
\rho_{B}(k,\eta) =\frac{k^3}{\pi^2} |b(k,\eta)|^2. 
\end{equation}
Thus, in terms of
\begin{equation}
r(\omega) = \frac{\rho_{B}(\omega,\eta)}{\rho_{\gamma},
(\eta)}
\end{equation}
and using Eqs.(\ref{mh})--(\ref{norm})
\begin{equation}
r_{B}(\omega,\eta_0) = C(\omega_1,\omega_{\rm r})
 {\cal T}(\omega)
\end{equation}
where 
\begin{eqnarray}
&&C(\omega_1,\omega_{\rm r}) = \zeta^2
 \frac{8}{\pi^2} \Gamma^2(5/2) 
\biggl(\frac{\omega_1}{\omega_{\rm r}}\biggr)^{4(\alpha +1)},
\nonumber\\
&& {\cal T}(\omega) = e^{ - \frac{\omega^2_{\rm G}}{\omega^2_{\sigma}}}
\bigl[ \bigl(\frac{\omega_{\vf}}{\omega_1} \bigr) 
\bigl( \frac{\omega_{\vf}}{\omega_{\rm m}}\bigr)^{\frac{1}{2}} 
\bigl( \frac{\omega_{\vf}}{\omega_{\rm c}}\bigr)^{\frac{3}{2}} \bigr]^{ 
2 \frac{\omega^2}{T^2_0}},  
\label{aux}
\end{eqnarray}
with $\omega_{\sigma} \sim \sqrt{s/\eta_0}$ and $T_0 \sim 10^{-13}$ GeV.
In Eq. (\ref{aux}) $\omega_{\phi}$, $\omega_{\rm m}$ and $\omega_{\rm c}$ 
are, respectively, the present values of $H_{\phi}$, $H_{\rm m}$ and 
$H_{\rm c}$. Using the notation of Eq. (\ref{om1}) we have that
\begin{eqnarray} 
&&\omega_{\rm m} \sim \epsilon^{-1/(\alpha + 1)}\, \xi^{1/2}\, \zeta^{(1- 
\alpha)/(2 \alpha + 2)}\, \omega_1,
\nonumber\\
&&\omega_{\rm c}  \sim \varphi^2 
\epsilon^{-1/(\alpha + 1)}\, \xi^{1/2}\, \zeta^{(1- 
\alpha)/(2 \alpha + 2)}\, \omega_1,
\nonumber\\
&&\omega_{\phi} \sim 
\xi^{7/6} \varphi^{2/3}  \epsilon^{-1/(\alpha + 1)}\, \zeta^{(1- 
\alpha)/(2 \alpha + 2)}\, \omega_1.
\label{def}
\end{eqnarray} 
All the frequencies are evaluated at the time $\eta_0$.
Using now the previous equations,
\begin{equation}
r_{B}(\omega_{\rm G}) \sim \epsilon^4 \zeta^{-2}.
\label{rb}
\end{equation}
where $\omega_{\rm G} \sim 10^{-14} $Hz is the 
present frequency corresponding to a Mpc scale.
This result should be confronted with typical values of $r_{\rm B}$ required
in order to explain galactic (and possibly inter-galactic) magnetic fields.
Before doing so let us discuss the main spectral features 
of the obtained results. The spectrum is flat for frequencies 
$\omega < \omega_{\sigma}$ where $\omega_{\sigma}$ is the 
present value of the magnetic diffusivity frequency. 
The frequency $\omega_{\sigma}$ 
roughly of the order of $10^{-3}$ Hz corresponding 
to a typical scale of the order of the astronomical unit. 
Thanks to the flatness of the spectrum, the 
constraints obtained on the parameters at the galactic scale will be 
preserved at even larger scales. The upper limit in the amplitude 
will be dictated by $\epsilon = 10^{-6}$ (as required 
by CMB observations) and $\zeta \simeq \epsilon$ (i.e. the case 
of instantaneous reheating). 
These features are summarized in Fig. \ref{F2}

\begin{figure}[htb]
    \centering
    \includegraphics[height=2.5in]{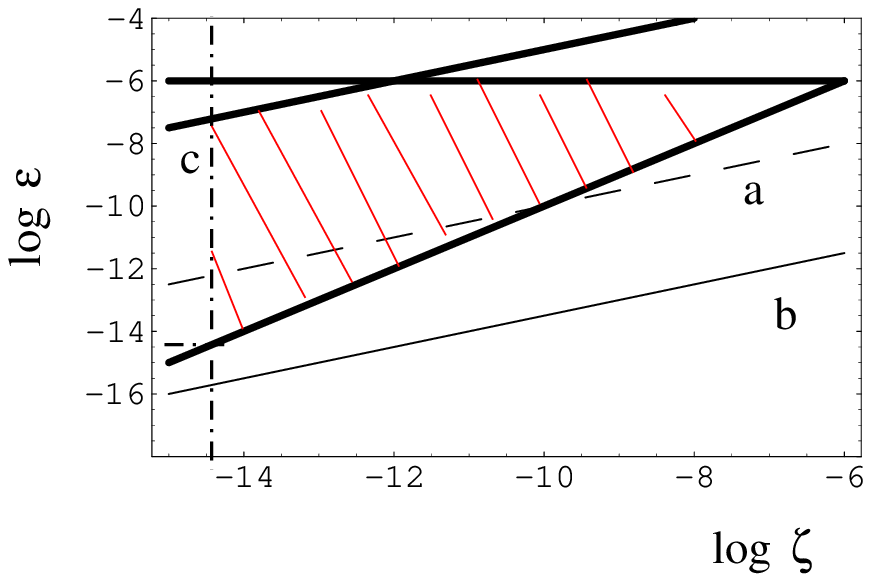}
    \caption{Magnetogenesis requirements 
are illustrated. The thin line {\bf b} corresponds to 
$r_{\rm B}(\omega_{\rm G}) \sim 10^{-34}$. The dashed line {\bf a} 
corresponds $r_{\rm B}(\omega_{\rm G}) \sim 10^{-20}$. The (horizontal)
thick line correspond to $\epsilon = 10^{-6}$. The (diagonal) thick line 
corresponds to $\epsilon \sim \zeta$. The thick line {\bf c} corresponds 
to the critical density bound which implies $r(\omega ) <1$. Since the 
spectrum is flat (or mildly increasing with the frequency), 
this bound is also satisfied at large scales.
Magnetogenesis is possible when the parameters lie in the shaded area  }
    \label{F1}
\end{figure}

In Fig. \ref{F1} the shaded area illustrates the region where 
magnetogenesis is possible.

To be consistent with inflationary production
of scalar and tensor fluctuations of the geometry $\epsilon < 10^{-6}$ 
should be imposed. Thus, in Fig. \ref{F1} the parameters 
should all lie below the (horizontal) thick line. Moreover, since 
$H_{\rm r} < H_{1}$, $\zeta < \epsilon$. Recall that $\zeta > 10^{-15}$  
in order not to affect the nucleosynthesis epoch [see the dot-dashed line 
in Fig. \ref{F1}]. This requirement comes about since $H_{\rm r} > m$ and
$\xi \geq 10^{-15}$. In \cite{dinng2} these considerations have been extended 
to more general evolutions of the gauge coupling.

\renewcommand{\theequation}{6.\arabic{equation}}
\setcounter{equation}{0}
\section{Faraday rotation of CMB}

If large scale magnetic fields are truly primordial they should 
be present prior to the decoupling epoch. Can these fields 
be ``detected'' in some way?  The effects of large scale magnetic fields
possibly present to the decoupling epoch are manifold. 
Large scale magnetic fields possibly present at the BBN epoch 
can have an impact on the light nuclei formation. By reversing 
the argument, the success of BBN can be used in order 
to bound the magnetic energy density possibly present at 
the time of formation of light nuclei.

These bounds are qualitatively  different from the ones previously quoted and
coming, alternatively, from homogeneity \cite{hom} and isotropy 
\cite{iso} of the background geometry at the BBN time.
As elaborated in slightly different frameworks
 through the years \cite{bbn1,bbn2,bbn3,bbn4,bbn5}, magnetic fields 
possibly present at the BBN epoch could have a twofold effect. 
On one hand they could 
enhance the rate of reactions (with an effect
proportional to $\alpha \rho_{B}$) 
and, on the other hand they could 
artificially increase the expansion rate (with an effect 
proportional to $\rho_{B}$). It turns out that 
the latter effect is probably the most relevant \cite{bbn4}. 
 In order to prevent the Universe 
from expanding too fast at the BBN epoch 
$\rho_{B} < 0.27 \rho_{\nu}$  where  $\rho_{\nu}$ is the 
energy density contributed by the standard three neutrinos for $T< 1$ MeV.

If magnetic fields are present at BBN there are no reasons why they should be 
absent at the decoupling epoch. Moreover, if quantum mechanical 
fluctuations of gauge fields are amplified by breaking Weyl invariance, then 
magnetic fields will be produced over different scales including scales 
larger than the Mpc. 

Large scale magnetic fields present at the decoupling epoch can have 
various consequences. For instance they can induce fluctuations in the 
CMB \cite{pbb1,GVD}, they can distort the Planckian spectrum of CMB 
\cite{jed}, 
they can distort the acoustic peaks of  CMB anisotropies \cite{rg} and 
they can also depolarize CMB \cite{FR}.

The polarization of the CMB
represents a very interesting observable which has been extensively
investigated in the past both from the theoretical \cite{1} and
experimental points of view \cite{2}. Forthcoming satellite missions
like PLANCK \cite{3} seem to be able to achieve a level
of sensitivity which will enrich decisively our experimental knowledge
of the CMB polarization with new direct measurements. 

If the background geometry of the universe is homogeneous but not
isotropic the CMB is naturally polarized \cite{1}. 
This phenomenon occurs, for example, in  Bianchi-type I models \cite{4}.
On the other hand if the background geometry is homogeneous and
isotropic (like in the Friedmann-Robertson-Walker [FRW] case) it seems very
reasonable that the CMB acquires a small degree of linear
polarization provided the radiation field has a non-vanishing
quadrupole component at the moment of last scattering \cite{5}.

Before decoupling photons, baryons and electrons form a unique fluid
which possesses only monopole and dipole moments, but not
quadrupole. Needless to say, in a homogeneous and isotropic model of
FRW type a possible source of linear polarization for the CMB becomes
efficient only at the decoupling and therefore a  small degree of linear
polarization seems a firmly established theoretical option  which will
be (hopefully) subjected to direct tests in the near future.
The discovery of a linearly polarized CMB could also have a
remarkable impact upon other (and related) areas of cosmology. 
Indeed the linear polarization of the CMB is a very promising
laboratory in order to directly probe the speculated existence of a
large scale magnetic field (coherent over the horizon size at the
decoupling) which might actually rotate (through the Faraday
effect \cite{rev1,rev2,rev3}) the polarization plane of the CMB. 

Consider, for
instance, a linearly polarized electromagnetic wave of physical 
frequency $\omega$
traveling along the $\hat{x}$ direction in a cold plasma of ions and
electrons together with  a magnetic field ($ \overline{B}$)
 oriented along an arbitrary direction ( which might coincide with
$\hat{x}$ in the simplest case). 
If we let the polarization vector at the origin ($x=y=z=0$, $t=0$) 
be directed along the $\hat{y}$ axis, after the
wave has traveled a length $\Delta x$, the corresponding angular shift
($\Delta\alpha$) in the polarization plane will be :
\begin{equation}
\Delta\alpha= f_{e} \frac{e}{2m}
\left(\frac{\omega_{pl}}{\omega}\right)^2 (\overline{B}\cdot\hat{x}) \Delta x
\label{Faraday1}
\end{equation}
(conventions: $\omega_{B} = e B/m $ is the Larmor frequency;
$\omega_{pl} = \sqrt{4\pi n_{e} e^2/m}$ is the plasma frequency $n_e$
is the electron density and $f_{e}$ is the ionization fraction ;
 we use everywhere natural units $\hbar = c = k_{B}=1$).
It is worth mentioning that the previous estimate of the Faraday
rotation angle $\Delta\alpha$ holds provided $\omega\gg\omega_{B}$ and
$\omega\gg\omega_{pl}$.
From Eq. (\ref{Faraday1})
by stochastically averaging over all the possible orientations 
of $\overline{B}$  and
by assuming that the last scattering surface is infinitely thin
(i.e. that  $\Delta x f_{e} n_{e} \simeq \sigma_{T}^{-1}$ where
$\sigma_{T}$ is the Thompson cross section) we
get an expression connecting the RMS of the rotation angle to the
magnitude of $\overline{B}$ at $t\simeq t_{dec}$
\begin{equation}
\langle(\Delta\alpha)^2 \rangle^{1/2} \simeq 1.6^{0} 
\left(\frac{B(t_{dec})}{B_{c}} \right)
\left(\frac{\omega_{M}}{\omega}\right)^2,~~~B_{c} =
10^{-3}~{\rm Gauss},~~~\omega_{M} \simeq  3\times10^{10}~Hz
\label{Faraday2}
\end{equation}
(in the previous equation we implicitly assumed that the frequency of
the incident electro-magnetic radiation is centered around the maximum
of the CMB).
We can easily argue from Eq. (\ref{Faraday2}) that if $B(t_{dec}) \gaq
B_c$ the expected rotation in the polarization plane of the CMB is
non negligible.
Even if we are not interested, at this level, in a precise estimate of
$\Delta\alpha$, we point out that more refined determinations of the
expected Faraday rotation signal (for an incident frequency
$\omega_{M}\sim 30~{\rm GHz}$) were recently carried out \cite{6b,6b2}
leading to a result fairly consistent with (\ref{Faraday1}).

Then, the statement is the following. {\em If} the CMB is linearly 
polarized and {\em if} a large scale magnetic field is 
present at the decoupling epoch, {\em then} the polarization plane of the 
CMB can be rotated \cite{FR}. The predictions of different 
models predicting the generation of large scale magnetic 
fields can then be confronted 
with the requirements coming from a possible detection of depolarization 
of the CMB \cite{FR}.
 
\renewcommand{\theequation}{7.\arabic{equation}}
\setcounter{equation}{0}
\section{Hypermagnetic fields, EWPT and relic gravitons}

If magnetic fields are generated over all physical scales compatible 
with the plasma dynamics they may have been present 
even before the BBN epoch, namely at the electroweak epoch. 
Some of these fields already decayed by today since 
they are washed out by simultaneous effects of finite magnetic and 
thermal diffusivities.  However, at the electroweak time 
these fields did not dissipate yet. When we talk about large scale 
magnetic fields at the electroweak scale we use, as unit, the electroweak 
time at the epoch of the phase transition. 

In this section the we will illustrate the idea 
that a classical hypermagnetic background 
can provide an explanation of the possible formation of the BAU 
giving rise, simultaneously, to stochastic gravitational 
waves background.

At small temperatures and small densities of different 
fermionic charges the $SU_{L}(2) \otimes U_{Y}(1)$  is 
broken down to the $U_{\rm em}(1)$ and the long range fields which can 
survive in the plasma are the ordinary magnetic fields. 
However, for sufficiently high temperatures the 
$SU_{L}(2) \otimes U_{Y}(1)$ is restored and
non-screened vector modes correspond to hypermagnetic fields. 
At the electroweak epoch the typical size of the horizon is of the 
order of $3$ cm . The typical diffusion scale is of the order of $10^{-9}$ cm.
Therefore, over roughly eight orders of magnitude hypermagnetic fields can 
be present in the plasma without being dissipated \cite{mg3}. 
The evolution of hypermagnetic fields can be obtained from the  anomalous 
magnetohydrodynamical (AMHD) 
equations. The AMHD equations generalize the treatment 
of plasma effects involving hypermagnetic fields to the case 
of finite fermionic density\cite{mg4}. 

Depending on their topology, hypermagnetic fields can have various consequences
\cite{mg3,mg4}. 
If the hypermagnetic flux lines have a trivial topology they can have an impact on 
the phase diagram of the electroweak phase transition \cite{h,h2}. 
If the topology of hypermagnetic fields is non trivial, hypermagnetic knots 
can be formed \cite{kn1} and, under specific conditions, the BAU
can be generated \cite{kn2}.

A classical hypermagnetic background in the symmetric phase of the EW theory can produce
interesting amounts of gravitational radiation  
 in a frequency range between $10^{-4}$ Hz
and the kHz. The lower tail falls into the LISA window while the 
higher tail falls in the VIRGO/LIGO window. For the hypermagnetic background 
required in order to seed the BAU the amplitude of the obtained GW 
can be even six orders of magnitude larger than the 
inflationary predictions. In this context, the mechanism of 
baryon asymmetry generation is connected with GW production \cite{kn1,kn2}.

\subsection{Hypermagnetic knots}

It is  possible to construct hypermagnetic knot configurations 
with finite energy and helicity which are localized in space and within 
typical distance scale  $L_{s}$. 
Let us consider in fact the following configuration
in spherical coordinates \cite{kn2}
\begin{eqnarray}
{\cal Y}_{r}({\cal R},\theta) &=& - \frac{2 B_0}{ \pi L_{s}}
\frac{\cos{\theta} }{\bigl[{\cal R}^2 +1\bigr]^2},
\nonumber\\
{\cal Y}_{\theta}({\cal R},\theta) &=& \frac{2 B_0}
{ \pi L_{s}} \frac{ \sin{\theta}}{\bigl[ {\cal
R}^2 + 1\bigr]^2},
\nonumber\\
{\cal Y}_{\phi}({\cal R},\theta) &=& - \frac{ 2 B_0}{ \pi L_{s}} \frac{ n
{\cal
R}\sin{\theta}}{\bigl[{\cal R}^2 + 1\bigr]^2},
\label{conf2}
\end{eqnarray}
where ${\cal R}= r/L_{s}$ is the rescaled radius and $B_{0}$ is some 
dimensionless amplitude and $n$ is just an integer number 
whose physical interpretation will become clear in a moment. 
The hypermagnetic field can be easily computed 
from the previous expression and it is 
\begin{eqnarray}
&&{\cal H}_{r}({\cal R},\theta) = - \frac{4 B_{0}}{\pi~ L_{s}^2}\frac{n
\cos{\theta}}{\bigl[  {\cal
R}^2 + 1\bigr]^2},
\nonumber\\
&&{\cal H}_{\theta}({\cal R}, \theta) = - \frac{4 B_{0}}{\pi~
L_{s}^2}\frac{{\cal R}^2 -1}{\bigl[
{\cal R}^2 + 1\bigr]^3}n \sin{\theta},
\nonumber\\
&&{\cal H}_{\phi}({\cal R}, \theta) = 
- \frac{8 B_0}{ \pi~ L_{s}^2}\frac{ {\cal R}
\sin{\theta}}{\bigl[
{\cal R}^2 + 1\bigr]^3}.
\label{knot}
\end{eqnarray}
The poloidal and toroidal components of $\vec{{\cal H}}$ can be usefully 
expressed as $\vec{{\cal H}}_{p} = 
{\cal H}_{r} \vec{e}_{r} + {\cal H}_{\theta} \vec{e}_{\theta} $ 
and $\vec{\cal H}_{t}= {\cal H}_{\phi} \vec{e}_{\phi}$.
The Chern-Simons number is finite and it is given by 
\begin{equation}
N_{CS} =\frac{g'^2}{32\pi^2}
\int_{V} \vec{{\cal Y}} \cdot \vec{{\cal H}}_{{\cal Y}} d^3 x=
\frac{g'^2}{32\pi^2} \int_{0}^{\infty}
\frac{ 8 n
B^2_0}{\pi^2} \frac{ {\cal R}^2 d {\cal R}}{\bigr[ {\cal R}^2 +
1\bigl]^4} = \frac{g'^2 n B^2_0}{32 \pi^2}.
\label{CS}
\end{equation}
We can also compute the total helicity of the configuration namely
\begin{equation}
\int_{V} \vec{{\cal H}}_{Y}
 \cdot \vec{\nabla} \times \vec{{\cal H}}_{Y} d^3 x=
\frac{256~B^2_0~n}{\pi L^2 } \int_{0}^{\infty} \frac{ {\cal R}^2 d
{\cal R}}{(1 + {\cal R}^2)^5} = \frac{5 B^2_0 n}{L_s^2}.
\label{helic}
\end{equation}
We can compute also the total energy of the field
\begin{equation}
E = \frac{1}{2}\int_{V} d^3 x |\vec{{\cal H}}_{Y}|^2 = \frac{B^2_0}{2~L_{s}}
(n^2 + 1).
\end{equation}
and we discover that it is proportional to $n^2$.
 This means that one way of increasing the total energy of
 the field is to increase the number of knots and twists in the flux lines.
We can also have some real space pictures of the core of the knot
 (i.e. ${\cal R} = r/L_{s}<1$).
This type of configurations can be also obtained by projecting a 
non-Abelian SU(2) (vacuum) gauge field on a fixed electromagnetic 
direction \cite{JPI} \footnote{ In order to interpret these solutions it is 
very iteresting to make use of the Clebsh decomposition. The 
implications of this  decomposition (beyond the hydrodynamical context, where 
it was originally discovered) have been recently discussed (see \cite{JPI2} 
and references therein). I thank R. Jackiw for interesting discussions about 
this point.}
These configurations have been also 
studied in \cite{ad1,ad2}.  In particular, in \cite{ad2}, the relaxation 
of HK has been investigated with a technique different from the one employed
 in \cite{kn1,kn2} but with similar results.

Topologically non-trivial 
configurations of the hypermagnetic flux lines lead to the formation 
of hypermagnetic knots (HK) whose decay 
might seed the Baryon Asymmetry of the Universe (BAU).
HK can be dynamically generated  provided
a topologically trivial (i.e. stochastic)
distribution of flux lines is already present 
in the symmetric phase of the electroweak (EW) theory \cite{kn1,kn2}. 
In spite of
the mechanism generating the HK, their typical size
must exceed the diffusivity length scale. 
In the  minimal standard model (MSM) (but not necessarily 
 in its supersymmetric extension) HK are washed out. 
A classical hypermagnetic background 
in the symmetric phase of the EW theory can 
produce interesting amounts of gravitational radiation.

The importance of the topological properties 
of long range (Abelian) hypercharge magnetic fields has been  
stressed in the  past \cite{vi,ru,ru2,ru3}.
In \cite{m1} it was argued that if the spectrum of hypermagnetic 
fields is dominated by parity non-invariant Chern-Simons 
(CS) condensates, the BAU could be the result of their decay. 
Most of the mechanisms
often invoked for the origin of large scale magnetic fields in the early 
Universe seem to imply the production of topologically trivial (i.e. 
stochastic) configurations of magnetic fields \cite{s1,s2,s3}.

\subsection{Hypermagnetic knots and BAU}

Suppose that the EW plasma is filled, for $T> T_{c}$ 
with topologically trivial hypermagnetic fields $\vec{\cal H}_{Y}$, 
which  can be physically pictured as a 
collection of flux tubes (closed because of the transversality 
of the field  lines)  evolving independently without 
breaking  or intersecting with each other. If the field
 distribution is topologically 
trivial (i.e. $\langle\vec{\cal H}_{Y} \cdot\vec{\nabla} 
\times\vec{\cal H}_{Y}\rangle =0$) parity is  a  good symmetry 
of the plasma and the field can be completely homogeneous. 
We name hypermagnetic knots those CS condensates carrying 
a non vanishing (averaged)  hypermagnetic helicity
(i.e.  $\langle\vec{\cal H}_{Y} \cdot\vec{\nabla} 
\times\vec{\cal H}_{Y}\rangle \neq 0$). 
If $\langle\vec{\cal H}_{Y} \cdot\vec{\nabla} 
\times\vec{\cal H}_{Y}\rangle \neq 0$  parity is  broken for scales 
comparable with the size of the HK,
 the flux lines are knotted and the field $\vec{{\cal H}}_{Y}$ 
cannot be completely homogeneous.  

In order to seed the BAU a network of HK should be present at high
temperatures \cite{mg3,mg4,m1}. In fact
for temperatures larger than $T_{c}$
 the fermionic number is stored both in HK 
and in real fermions.  For $T<T_{c}$, 
the HK should release real fermions 
since the ordinary magnetic fields (present {\em after} EW 
symmetry breaking) do not carry fermionic number.
If the EWPT is strongly first order the decay of the HK 
can offer some seeds for the BAU generation \cite{m1}.
This last condition can be met in the 
minimal supersymmetric standard model (MSSM) \cite{PT,mssm,mssm1,mssm2}.

Under these hypotheses the integration of the $U(1)_{Y}$ 
anomaly equation \cite{m1}
gives the CS number density carried by the HK
which is in turn related to the density of baryonic number $n_{B}$
for the case of $n_{f}$ fermionic generations.
\begin{equation}
\frac{n_{B}}{s}(t_{c})=
\frac{\alpha'}{2\pi\sigma_c}\frac{n_f}{s}
\frac{\langle{\vec{{\cal H}}}_{Y}\cdot \vec{\nabla}\times
{\vec{{\cal H}}}_{Y}\rangle}{\Gamma + \Gamma_{{\cal H}}}
\frac{M_{0}\Gamma}{T^2_c},~~\alpha' = \frac{g'^2}{4\pi}
\label{BAU}
\end{equation}
($g'$ is the $U(1)_{Y}$ coupling and $s = (2/45) \pi^2 N_{eff}T^3$ 
is the entropy density; $N_{eff}$, at $T_{c}$,
 is $106.75$ in the MSM;
$M_{0}= M_{P}/1.66 \sqrt{N_eff} \simeq 7.1 \times 10^{17} {\rm GeV}$).
In Eq. (\ref{BAU}) $\Gamma$ is the perturbative rate of the 
right electron chirality 
flip processes  (i.e. 
scattering of right electrons with the Higgs and gauge bosons and with 
the top quarks because of their large Yukawa coupling) which 
are the slowest reactions in the plasma and 
\begin{equation}
\Gamma_{{\cal H}} = \frac{783}{22} \frac{\alpha'^2}{\sigma_{c} \pi^2} 
\frac{|\vec{{\cal H}}_{Y}|^2}{T_{c}^2}
\end{equation}
is the rate of right electron dilution induced by the presence of a
 hypermagnetic field. In the MSM we have that 
$\Gamma < \Gamma_{{\cal H}}$ \cite{fl} 
whereas in the MSSM $\Gamma$ can naturally 
be larger than $\Gamma_{{\cal H}}$. 
Unfortunately, in the MSM 
a hypermagnetic field can modify the phase diagram of the phase transition 
but cannot make the phase transition strongly first order for large masses of
the Higgs boson \cite {h}. 
Therefore, we will concentrate on the case $\Gamma > \Gamma_{\cal H}$ and we
 will show that in the opposite limit the BAU will be anyway small 
even if some (presently unknown) mechanism would make the EWPT strongly 
first order in the MSM.

HK can be dynamically generated \cite{kn1,kn2}.
Gauge-invariance 
and transversality of the magnetic fields suggest
 that perhaps
 the only way of producing  $\langle\vec{\cal H}_{Y} \cdot\vec{\nabla} 
\times\vec{\cal H}_{Y}\rangle \neq 0$ is to 
postulate, a time-dependent interaction between the two (physical) 
 polarizations of the hypercharge field $Y_{\alpha}$.
Having defined the Abelian field strength  $Y_{\alpha\beta} = 
\nabla_{[\alpha} Y_{\beta]}$ and its dual $\tilde{Y}_{\alpha\beta}$ such an 
 interaction can be described, in curved space, by the 
Lagrangian density
\begin{equation}
L_{eff}= \sqrt{-g} \biggl[ 
-\frac{1}{4}Y_{\alpha\beta} Y^{\alpha\beta} + 
c\frac{\psi}{4 M}
Y_{\alpha\beta}\tilde{Y}^{\alpha\beta}\biggr].
\label{action}
\end{equation}
where $g_{\mu\nu}$ is the metric tensor and $g$ its determinant, 
$c$ is the coupling  constant and $M$ is a typical scale.
This  interaction 
is plausible  if the $U(1)_{Y}$ anomaly is coupled, 
(in the symmetric phase of the EW theory ) to  dynamical 
pseudoscalar particles $\psi$ (like the axial Higgs of the MSSM).
Thanks to the presence of pseudoscalar particles, 
the two polarizations of $\vec{{\cal H}}_{Y}$ 
evolve in a slightly different way
producing, ultimately, inhomogeneous HK.

Suppose that 
an inflationary phase with $a(\tau) \sim \tau^{-1}$  is continuously 
matched, at the transition time $\tau_1$, to a radiation dominated 
phase where $a(\tau)\sim \tau$. Consider then a massive pseudoscalar 
field $\psi$ which oscillates during the last  stages 
of the inflationary evolution
with typical amplitude $\psi_0 \sim M$.
As a result of the inflationary evolution
 $|\vec{\nabla}\psi| \ll \psi'$. Consequently, the 
phase of $\psi$  can  get frozen. 
Provided  the 
pseudoscalar mass $m$ is larger than the inflationary curvature  scale 
$H_{i}\sim {\rm const.}$, the $\psi$ oscillations are converted, 
at the end of the
 quasi-de Sitter stage, in a net helicity arising as a result 
of the different evolution of the two (circularly polarized) vector potentials
\begin{eqnarray}
&& {Y}_{\pm}'' + \sigma Y'_{\pm}+
\omega_{\pm}^2 {Y}_{\pm} =0,~~\vec{H}_{Y} = \vec{\nabla} 
\times \vec{Y}
\label{Y}\\
&& \omega_{\pm}^2 = k^2 \mp k \frac{c}{M} a \dot{\psi}
\end{eqnarray}
(where we denoted with $\vec{H}_{Y} = a^2 {\vec{\cal H}}_{Y}$  the 
curved space fields and with $\sigma= \sigma_c a$ the rescaled 
hyperconductivity; the prime denotes derivation with respect to 
conformal time $\tau$ whereas the over-dot denotes differentiation 
with respect to cosmic time $t$). 

Since $\omega_{+}\neq \omega_{-}$ the helicity 
gets amplified according to Eq. (\ref{Y}). 
There are two important points to stress in this context. 
First of all the plasma effects as well as the finite density effects 
are important. This means, in practical terms, that the dissipation scales 
of the problem should be borne in mind. This has not always 
been done. The second point is related to the first. By treating, consistently, 
the plasma and finite density effects in the context of AMHD \cite{mg3,kn1,kn2}, 
one realizes that the pseudoscalar coupling of $\psi$ together 
with the coupling of the coupling to the chemical potential are not 
sufficient in order to seed the BAU {\em unless} some 
hypermagnetic background is already present. 
In other words the scenario which leads to the 
generation of the BAU is the following. 
Correlation in the hypermagnetic fields are generated 
outside the horizon durin inflation by direct breaking 
of conformal invariance. An example in this direction 
has been given in the 
previous Section. Then hypermagnetic fields 
re-entering at the electroweak epoch will participate 
in the dynamics and, in particular, they will 
feel the effect of the anomalous coupling either to $\psi$ or to 
the chemical potential. The effect of the anomalous coupling 
will not be to amplify the hypermagnetic background. The 
anomalous coupling will make the topology of the hypermagnetic 
flux lines non-trivial. So the statement is that 
{\em if} conformal invariance is broken and, {\em if} hypermagnetic 
have anomalous couplings, {\em then} a BAU $\gaq 10^{-10}$ can be achieved
without spoiling the standard cosmological evolution \cite{kn1,kn2}.
It is worth mentioning that this type of scenario may be motivated by 
the low energy string effective action where, by supersymmetry, 
Kalb-Ramond axions and dilatons are coupled, respectively, to 
the anomaly and to the gauge kinetic term \cite{kr}.
It is interesting to notice that, in this scenario, the value 
of the BAU is determined by various particle physics parameters but also 
by the ratio of the hypermagnetic energy density over the 
energy density sitting in radiation during 
the electroweak epoch, namely, using the language 
of the previous section \cite{kn1,kn2}, 
\begin{equation}
\frac{n_{B}}{s} \propto r.
\end{equation}
In order to get a sizable BAU $r$ should be at least $10^{-3}$ 
if the anomalous coupling operates during a radiation 
phase. The value of $r$ coupld be smaller in models where 
the anomalous coupling is relevant during a low scale inflationary phase \cite{kn2}.

\subsection{Hypermagnetic fields and relic gravitons}    

If
a hypermagnetic background is present for $T> T_c$, then, as discussed
in \cite{mmm} in the context of ordinary MHD,  the energy momentum tensor 
will acquire a small anisotropic component which will source the evolution 
equation of the tensor fluctuations $h_{\mu\nu}$ of the metric $g_{\mu\nu}$: 
\begin{equation}
h_{ij}'' + 2 {\cal H} h_{ij}' - \nabla^2 h_{ij} = - 16 \pi G
\tau^{(T)}_{ij}.
\label{GWeq}
\end{equation}
where $\tau^{(T)}_{ij}$ is the {\em tensor} component of the 
{\em energy-momentum tensor} \cite{mmm} 
of the hypermagnetic fields. Suppose now, as assumed in \cite{h} that 
$|\vec{{\cal H}}|$ has constant amplitude and that it is also 
homogeneous. Then 
as argued in  \cite{griru} we can easily deduce 
the critical fraction of energy density  present today in relic gravitons 
of EW origin 
\begin{equation}
\Omega_{\rm gw}(t_0) = \frac{\rho_{\rm gw}}{\rho_c} 
\simeq z^{-1}_{{\rm eq}}
r^2,~~\rho_{c}(T_{c})\simeq N_{\rm eff} T^4_{c }
\end{equation}
($z_{\rm eq}=6000$ is the redshift from the time of matter-radiation,
 equality to the present time 
$t_0$). Because of the structure of the AMHD equations, stable 
hypermagnetic fields will be present not only for 
$\omega_{\rm ew}\sim k_{\rm ew}/a$ but 
for all the range $\omega_{{\rm ew}} <\omega< \omega_{\sigma}$ where 
$\omega_{\sigma}$ is the diffusivity frequancy. Let us assume, 
 for instance, that $T_{c} \sim 100 $ GeV and $N_{eff} = 106.75$. 
Then, the (present) values of 
$\omega_{\rm ew}$ is 
\begin{equation}
\omega_{\rm ew } (t_0) \simeq 2.01 \times 10^{-7} \biggl( \frac{T_{c}}{1 {\rm GeV}} \biggr) 
\biggl(\frac{ N_{\rm eff}}{100} \biggr)^{1/6} {\rm Hz} .
\end{equation}
Thus, $\omega_{\sigma}(t_0) \sim 10^{8} \omega_{\rm ew} $. Suppose now that 
$T_{c} \sim 100$ GeV; than we will have that $\omega_{\rm ew}(t_0) \sim 10^{-5}$ Hz. 
Suppose now, as assumed in \cite{h}, that 
\begin{equation}
|\vec{{\cal H}}|/T_{c}^2 \gaq 0.3.
\end{equation} 
This requirement imposes $ r \simeq 0.1$--$0.001$ and, consequently, 
\begin{equation}
h_0^2 \Omega_{\rm GW} \simeq 10^{-7} - 10^{-8}.
\end{equation}
Notice that this signal would occurr in a (present) frequency 
range between $10^{-5}$ and $10^{3}$ Hz. This signal 
satisfies the presently available phenomenological 
bounds on the graviton backgrounds of primordial origin.
The pulsar timing bound ( which applies for present 
frequencies $\omega_{P} \sim 10^{-8}$ Hz and implies 
$h_0^2 \Omega_{\rm GW} \leq 10^{-8}$) is automatically satisfied
since our hypermagnetic background is defined for $10^{-5} {\rm Hz} 
\leq \omega \leq 10^{3} {\rm Hz}$. The large scale bounds would imply 
$h_0^2 \Omega_{\rm GW} < 7 \times 10^{-11}$ but a at much lower frequency 
(i.e. $10^{-18 }$ Hz). The signal discussed here is completely 
absent for frequencies $\omega < \omega_{\rm ew}$. Notice that 
this signal is clearly distinguishable from other stochastic 
backgrounds occurring at much higher frequencies (GHz region) 
like the ones predicted by quintessential inflation \cite{gw1}.
It is equally distinguishable from signals due to 
pre-big-bang cosmology (mainly in the window of
ground based interferometers \cite{gw2}).
The frequency of operation of the interferometric devices 
(VIRGO/LIGO) is located between few Hz and 10 kHz \cite{gw2}.
 The frequency of operation 
of LISA is well below the Hz. Hence, the background described here 
lies in the LISA window \cite{kn1,kn2}. In this model the signal 
can be located both in the LISA window and in the VIRGO/LIGO window
due to the hierarchy between the hypermagnetic diffusivity scale and the 
horizon scale at the phase transition.

\section{Concluding remarks}

Fifty years after the first seminal speculations on the existence 
of large scale magnetic fields there are still various dark corners. 
It is crucial to improve the present evidence on large scale magnetic 
fields inside clusters. The existence of 
 $\mu $ Gauss magnetic fields  inside clusters would be interesting both 
for magnetogenesis models.The ongoing progress in x-ray astronomy
(driven by satellite missions) seems to be rather promising in order 
to improve our knowledge of the density of charged particles inside 
the clusters. Even more important 
would be to obtain observational evidence of magnetic fields right 
outside the Milky Way. At the moment there is no conclusive 
evidence of the fact that these fields should be as small as the n Gauss. 
They can be indeed larger. Another reasonable question often asked 
discussing large  scale magnetic fields concerns the possibility of their 
primordial origin. In this direction an interesting observable 
would certainly be the polarization of the CMB. Moreover, large scale 
magnetic fields present prior to BBN could give rise to an 
interesting stochastic GW background in the LISA window.

The detection of large scale magnetic fields inside clusters 
puts pressure on many models of magnetogenesis. Few  models 
predict magnetic fields of reasonable magnitude in this 
respect. A more complete theoretical understanding of the dynamo 
mechanism in clusters is certainly needed. 

As far as the problem of primordial magnetogenesis is concerned.
Few remarks are in order. Causal mechanisms may work at least 
for the galactic magnetic fields (where the understanding 
of the dynamo action is more convincing) {\em provided} 
some inverse cascade is realized.As far as inflationary mechanisms 
are concerned a crucial problem is to find a consistent way in order to break 
conformal invariance. If gauge couplings are dynamical,  conformal
invariance may be plausibly broken in the early stages of the 
inflationary evolution.

\Acknowledgements
The author wishes to thank M. E. Shaposhnikov 
for inspiring collaboration. Interesting discussions with 
M. Laine, H. Kurki-Suonio, E. Sihvola, K. Jedamzik are also 
acknowledged. Finally the author wishes to thank K. Enqvist, K. Kajantie, 
K. Kainulainen and M. Ross for the kind invitation to the COSMO-01 meeting.

\end{document}